# Symmetry analysis of magnetoelectric effects in perovskite – based multiferroics

Zukhra Gareeva [1,2,*], Anatoly Zvezdin [3], Konstantin Zvezdin[3] and Xiang Ming Chen [4]

1   Institute of Molecule and Crystal Physics, Subdivision of the Ufa Federal Research Centre of the Russian Academy of Sciences, 450075, Ufa, Russia; zukhragzv@yandex.ru,
2   Bashkir state University, 450076, Ufa, Russia,
3   Prokhorov General Physics Institute of the Russian Academy of Sciences, 119991, Moscow, Russia, zvezdin.ak@phystech.edu, zvezdin.ka@phystech.edu,
4   Laboratory of Dielectric Materials, School of Materials Science and Engineering, Zhejiang University, Zheda Road 38, 310027, Hangzhou, China, xmchen59@zju.edu.cn
*   Correspondence: zukhragzv@yandex.ru; Tel.: +7(917)8049560

**Abstract:** In this article, we perform the symmetry analysis of perovskite – based multiferroics: bismuth ferrite ($BiFeO_3$) - like, orthochromites ($RCrO_3$), and Ruddlesden – Popper perovskites ($Ca_3Mn_2O_7$ - like), being the typical representatives of multiferroics of the trigonal, rhombic, and tetragonal crystal families and explore the effect of crystallographic distortions on magnetoelectric properties. We determine the principal order parameters for each of the considered structures and obtain their invariant combinations consistent with the particular symmetry. This approach allowed us to analyze the features of the magnetoelectric effect observed during structural phase transitions in $Bi_xR_{1-x}FeO_3$ compounds and show that the rare- earth sublattice gives an impact into the linear magnetoelectric effect allowed by the symmetry of the new structure. It is shown that the magnetoelectric properties of orthochromites are attributed to the couplings between the magnetic and electric dipole moments arising near $Cr^{3+}$ ions due to distortions linked with rotations and deformations of the $CrO_6$ octahedra. For the first time, such symmetry consideration was implemented in the analysis of the Ruddlesden – Popper structures, which demonstrates the possibility of realizing the magnetoelectric effect in the Ruddlesden – Popper phases containing magnetically active cations and allows to estimate conditions required for its optimization.

**Keywords:** keyword multiferroics; crystal lattices; symmetry; perovskites; magnetoelectric effect

## 1. Introduction

Multiferroic, multifunctional materials, attract enormous research activity due to the prospects of their implementation in science in technology. The collective state parameters attainable in multiferroics (MFs) are preferable for advanced spintronic devices based on the use of combinatorial functions due to their high efficiency, low power consumption and adaptability [1].

Despite the variety of currently known magnetoelectric materials, perovskite – like structures remain the most researched MFs compounds. The family of perovskites stands out for its diversity due to intrinsic instability of the cubic parent perovskite phase [2]. Crystallographic distortions owed to substitution of various cations at A and B positions can lead to emergence of ferroelectric and MF properties due to the hybridization of ion orbitals, tilt of oxygen octahedra ($BO_6$), doping of various cations, and other factors [3]. $ABO_3$ perovskites can produce a variety of MFs differing in the type of i) ferroelectric ordering (A– or B –driven ferroelectricity, as in the cases $EuTiO_3$ [4] and $BiFeO_3$ [5] ii) magnetic orderings (although most perovskite MFs are G – type antiferromagnets (AFM), canted ferromagnetism and a combination of ferromagnetic and AFM phases are also possible) [6,7]; iii) the mechanism of magnetoelectric coupling, which can be associated with magnetoelectrically active d-ions or f-ions ($RFeO_3$)[8], the Dzyaloshinskii – Moriya interaction ($BiFeO_3$) [5] , spin-orbit coupling (BiMnTe) [9], antiferroelectric orderings ($RCrO_3$))[10] and iiii) temperatures of ferroelectric and magnetic orderings which remain low enough for most perovskite-based MFs.

Quite recently MF community turn to hybrid layered perovskite structures, considering them as the perspective candidates for high – temperature MFs. We refer to layered structures composed of an *n*-layered perovskite with a single-layer spacer in – between. Dependent on the kind of single layer spacer one can distinguish the Ruddlesden–Popper (RP) $A'_2(A_{n-1}B_nO_{3n+1})$, Dion–Jacobson (DJ) $A'(A_{n-1}B_nO_{3n+1})$ and Aurivillius $Bi_2O_2(A_{n-1}B_nO_{3n+1})$ structures [11,12]. When $n = \infty$, the layered perovskite transforms into a classical perovskite.

Among layered perovskites, the most widespread and studied compounds are the RP structures. They received their name in honor of the chemists R. Ruddlesden and P. Popper, who in 1957 synthesized the series of $Sr_2TiO_4$, $Sr_3Ti_2O_7$ and $Sr_4Ti_3O_{10}$ phases, forming a class of complex oxides $S_{m+1}Ti_nO_{3n+1}$ [13]. Since then, the class of RP structures has expanded significantly, at present it includes various compounds with the general formula $A'_2(A_{n-1}B_nO_{3n+1})$, where $A'$=La, Sr, Li,…, A=Sr, Ca, K,…., B=Ti, Fe, Mn, Nb…. The advantages of RP - type perovskites are related with long-term chemical stability, the activity of oxygen reduction reaction, and the possibility of realizing non-trivial magnetic and electrical properties [11–17]. Due to the activity of the reaction with oxygen, RP structures find application in solid oxide fuel cells [14]. As for magnetoelectricity, interest in layered perovskites is associated with the discoveries of the colossal magnetoresistance effect in the early 2000s in RP structures with magnetoactive cations ($Ca_3Mn_2O_7$, ….) [14–19] and ferroelectricity at room temperatures, found in $Ca_{3-x}Sr_xTi_2O_7$ in 2015 [20] and then in $Sr_3M_2O_7$-based oxides and other RP phases [21]. Currently, RP structures are considered promising candidates for high – temperature MFs. The crystal structure, instabilities, and structural phase transitions occurring in RP oxides are fairly well known; however, a number of questions concerning ferroelectricity, magnetic ordering and their coupling is open for discussion.

The focus of this manuscript is studying of the magnetoelectric effect in $Bi_{1-x}R_xFeO_3$, $RCrO_3$, $Ca_3(Ti_{1-x}Mn_x)_2O_7$ multiferroics (MFs). We aim to develop a unified approach to the analysis of the perovskite – like multiferroics based on the symmetry of the material, taking crystallographic distortions as the primary order parameters.

It is valuable to note that the classification of distortions and the search for recipes for effective magnetoelectric couplings in MFs with a perovskite structure is a long-standing and, at the same time, "hot" problem. We refer to several reviews, references therein and original papers that discussed various classification schemes for perovskite distortions (Glazer, Aleksandrov & Bartolome, and Fennie) and group – theoretical approaches, including their implementation in online tools (in particular, ones on the Bilbao Crystallographic Server) for identification of active order parameters, their possible couplings and invariant polynomials [22-29]. However, distortion classification schemes mainly deal with $BO_6$ octahedrons and do not account for polar distortions, which give significant impact into ferroelectric and MF properties, in addition, they also have limitations for specific systems. Despite the power of the software tools, they have not yet been applied to each MF system, for example, the details of the magnetoelectric coupling in MFs with a variable concentration of rare – earth ions have yet not been investigated, the same statement is also applied to the R-P structures.

So, in our research we appeal to group theoretical analysis, which is an elegant and effective tool for studying the properties of crystals with a complex magnetic structure, to classify the distortive, ferroelectric and magnetic orderings in several classes of perovskite MFs. For this purpose, we examine typical MF structures with the perovskite parent phase: $BiFeO_3$, $RCrO_3$, $Ca_3Mn_2O_7$, in which small distortions of the crystallographic lattice lead to systems of different symmetry (trigonal, orthorhombic, and tetragonal). Using the methods of theoretical group analysis, we consider how the symmetry manifests itself in their magnetoelectric properties. We determine magnetic and structural modes; classify them according to the irreducible representations of the corresponding symmetry group and calculate the coefficients of magnetoelectric couplings.

## 2. Materials and methods

In this section, we analyze the crystal structure, ferroelectric and magnetic properties of representatives of the perovskite – based multiferroics and, using the methods of the theoretical group analysis, determine the features of their magnetoelectric properties.

### 2.1. Multiferroic BiFeO3

We start our consideration with the well – known multiferroic $BiFeO_3$ with high – temperatures of ferroelectric and magnetic orderings ($T_C$=1083 K, $T_N$=643 K). $BiFeO_3$ crystallizes in the non – polar symmetry group R3c and is a ferroelectric material. The high – symmetry perovskite group is reduced to R3c due to three types of distortions: i) relative displacement of Bi and Fe ions along <111> axis, ii) deformations of oxygen octahedrons and iii) counterrotation of oxygen octahedrons around Fe ions. The displacements of Bi and Fe ions from their centrosymmetric positions lead to spontaneous electric polarization $P_s$ directed along one of the <111> crystallographic axes [30]. The magnitude of polarization has been the subject of controversy for a while. As was reported in Ref. [5,20,31–35] spontaneous polarization in BiFeO3 crystal and films can attain the values varying from 6 to 150 μC/cm². Electric polarization is sufficiently low $P_s$ ~ 6- 9 μC/cm² in single crystals and ceramic samples [34,35], the large values of $P_s$, being around 100 μC/cm², are achieved under epitaxial strain in the films [5,36–38], they can also be related to the supertetragonal phase induced by the strain.

The rigorous estimation of electronic polarization requires implementation of the quantum mechanics including electronic structure methods with correspondence to experimentally measurable observables [39]. In the frame of these approach in Ref. [40] electric polarization in perovskite BiFeO3 was calculated as a function of percentage distortion from the high symmetry non-polar structure to the ground state R3c structure, which gives the value 95.0 μC cm⁻². However, as was shown in Ref. [30] the similar results can be obtained with the use of the point charge model where the electric polarization is represented through the atomic displacements of $Bi^{3+}$ and $O^{2-}$ ions

$$P = \frac{4e}{V}\left(3\zeta_R - 4\zeta_{O1} - 2\zeta_{O3}\right) \quad (1)$$

where $e$ is the elementary charge, $V$ is the unit cell volume, $\zeta_R$ is the displacement of Bi ions corresponding to (i) distortion, $\zeta_{O1}$, $\zeta_{O3}$ are the displacements of oxygen ions corresponding to (ii) – distortion.

In magnetic relation BiFeO3 is weak ferromagnet with G - type antiferromagnetic (AFM) ordering characterized by AFM vector $\boldsymbol{L} = \frac{1}{V}\sum_{i=1}^{6}(-1)^{i}\boldsymbol{\mu}_i$, where $\mu_i$ are the magnetic moments of six Fe ions in a unit cell. Canting of spins is related with rotations of oxygen octahedrons FeO6, determined by the vector Ω, the distortions of (iii) – type [41], which give impact into weak ferromagnetic vector determined as

$$\boldsymbol{M} = \frac{V_0 a^2}{6J}\boldsymbol{\Omega}\times\boldsymbol{L} + \frac{V_0 a^2}{6J}\sum_{n=1}^{6}\left(\boldsymbol{n}_n\times\boldsymbol{\zeta}_{O_n}\right) \quad (2)$$

where $V_0$ is constant, $a$ is the lattice parameter, $J$ is the exchange constant, $n_n$ is the direction vector, oriented along one of <100> crystallographic axes. Here we used an assumption that the Heisenberg exchange constant remains unchanged, which is commonly used approximation for multiferroic BiFeO3, in particular, the deviation (non – collinearity) of the nearest neighbor spins is accounted by the antisymmetric Dzyaloshinskii – Moriya exchange interaction [42,43]. AFM spin arrangement is superimposed with spatially modulated magnetic structure with a large period λ = (620± 20 Å) incommensurate with the lattice parameter [44]. The magnetic moments of Fe ions retain their local mutually antiferromagnetic G-type orientation and rotate along the propagation direction of the modulation in the plane perpendicular to the basal crystal plane.

To explore magnetoelectric effect (MEE) in BiFeO3 we appeal to symmetry analysis. Knowledge of crystal structure and symmetry is of primary importance for this approach. To consider the properties of BiFeO3 the space group $R\bar{3}c$ is used to be taken as a "parent" symmetry phase [45]. $R\bar{3}c$ differs from the space symmetry group R3c of BiFeO3 single crystals only by the presence of a polar vector P = (0, 0, $P_s$). The symmetry analysis of the magnetoelectric properties of BiFeO3 has been performed in Refs. [45–47]. Here we focus our attention on the magnetoelectric tensor $\alpha_{ij}$, which links the electric polarization and the magnetic field

$$P_i = \alpha_{ij} H_j. \quad (3)$$

As was shown in Ref. [47] for BiFeO3 crystals, the tensor $\alpha_{ij}$ can be expressed through the components of the AFM vector $L$

$$\alpha_{ij}^{BiFeO_3} = \begin{vmatrix} -a_1 L_x & a_4 L_z + a_1 L_y & a_2 L_y \\ a_1 L_y - a_4 L_z & a_1 L_x & -a_2 L_x \\ -a_3 L_y & a_3 L_x & 0 \end{vmatrix} \quad (4)$$

As seen from eqs. (3), (4), the symmetry allows the linear magnetoelectric effect, however, due to the presence of an incommensurate space modulated structure, it is not observed in BiFeO3. So, when the spin cycloidal structure is destroyed, the linear magnetoelectric effect is restored. It can be achieved by applying a strong magnetic field, pressure or doping BiFeO3 with rare earth ions. In recent years, a series of experimental studies has been conducted, indicating the enhancement of the magnetoelectric effect in BiFeO$_3$ multilayers and composites.

Here, we discuss the replacement of various rare-earth (RE) ions at the A-positions as a way to improve the MF properties of BiFeO$_3$, a route that was paved in the 1990s [30,45–50] and current research confirm its efficiency [6,51]. As was shown in Ref. [51], La-substitution in Bi$_{1-x}$La$_x$FeO$_3$ (x = 0.08~0.22) ceramics induces continuous structural evolution from *R3c* to *Pna*2$_1$ and finally *Pbnm* phase. To understand the properties of Bi$_{1-x}$R$_x$FeO$_3$ structures (R=La, Gd, Dy, …) it is reasonable to introduce the AFM vector (*l*), which characterizes the magnetic ordering of the rare earth sublattice and consider *Pbnm* space symmetry group as the parent phase in Bi$_{1-x}$R$_x$FeO$_3$ structure, keeping in mind structural phase transitions *Pbnm* → *Pna*2$_1$. The space group *Pbnm* contains eight irreducible representations, their matrix representations are given in the columns corresponding to the generating symmetry elements (Table 1). The vector components of magnetic field *H*, electric polarization *P*, magnetization *M*, antiferromagnetic vectors *L* and *l* form the irreducible representations and are placed in the last column in Table 1 according to their transformation properties. Table 1 allows to determine transformation properties of AFM vector components, exchange – coupled structures, and magnetoelectric invariants, giving contribution to the magnetoelectric energy of Bi$_{1-x}$R$_x$FeO$_3$ structure

$$\Phi_{me}^{B_{1-x}R_xFeO_3} = P_z\left(\gamma_1\left(l_zM_y - l_yM_z\right) + \gamma_2\left(l_xL_y - l_yL_x\right)\right) + P_x\left(\gamma_{xyx}\left(l_yM_x - l_xM_y\right) + \gamma_2\left(l_yL_z - l_zL_y\right)\right) + P_y\left(\gamma_3l_xM_x + \gamma_2l_xL_z + \gamma_4\left(l_yM_y + L_y^2\right)\right) \quad (5)$$

where $\gamma_i$ are the magnetoelectric coefficients,
$\gamma_1 = \gamma_{zzy}, \gamma_2 = \gamma_{xyz}, \gamma_3 = \gamma_{yxx}, \gamma_4 = \gamma_{yyy}$

**Table 1.** Irreducible representations of the *Pbnm* symmetry group.

| $\Gamma_i$ | $\bar{1}$ | $2_x$ | $2_y$ | $2_z$ | Order parameters, magnetic and electric fields |
|---|---|---|---|---|---|
| $\Gamma_1$ | 1 | 1 | 1 | 1 | $L_y$ |
| $\Gamma_2$ | 1 | 1 | −1 | −1 | $M_x$  $L_z$  $H_x$ |
| $\Gamma_3$ | 1 | −1 | 1 | −1 | $M_y$  $H_y$ |
| $\Gamma_4$ | 1 | −1 | −1 | 1 | $M_z$  $L_x$  $H_z$ |
| $\Gamma_5$ | −1 | 1 | 1 | 1 | $l_y$ |
| $\Gamma_6$ | −1 | 1 | −1 | −1 | $l_x$  $P_z$  $E_z$ |
| $\Gamma_7$ | −1 | −1 | 1 | −1 | $P_y$  $E_y$ |
| $\Gamma_8$ | −1 | −1 | −1 | 1 | $l_z$  $P_x$  $E_x$ |

Magnetoelectric coupling coefficient for the Bi$_{1-x}$R$_x$FeO$_3$ structures can be determined as follows

$$\alpha_{ij} = \begin{vmatrix} -a_1l_y & a_{12}l_x & a_{13}\left(l_xL_z + l_zL_x\right) \\ a_{12}l_x & a_2l_y & a_{23}l_z \\ a_{13}\left(l_xL_z + l_zL_x\right) & -a_{23}l_z & a_{23}l_y \end{vmatrix} \approx \begin{vmatrix} -a_1l_y & a_{12}l_x & 0 \\ a_{12}l_x & a_2l_y & a_{23}l_z \\ 0 & -a_{23}l_z & a_{23}l_y \end{vmatrix} \quad (6)$$

In the case of Bi$_{1-x}$R$_x$FeO$_3$, rare earth elements can give a significant impact on magnetoelectric properties. As seen from eq. (5), $\alpha_{ij}$ depends on the components of the AFM vector of the rare earth sublattice (for the case of non –zero magnetization on the R ion site), and it is possible to distinguish a linear MEE related to rare earth elements, and a quadratic MEE emerging due to cross-coupling of the AFM vectors of iron and rare – earth sublattices.

*2.2. Magnetoelectric properties of RCrO₃*

In this section we consider MEE in the rare earth orthochromites (RCrO$_3$), focusing on the relation of magnetoelectric properties with crystallographic distortions. The crystal structure of RCrO$_3$ belongs to the space symmetry group *Pbnm* [52–54]. Its orthorombically distorted perovskite unit cell contains 4 RCrO$_3$ molecules (Fig.3). 4 Cr$^{3+}$ and 4 R$^{3+}$ magnetic ions (in the case of R = Rare Earth Ion) are located in the local positions differing by the symmetry of O$^{2-}$ environments. The *d* – ions (M$^{3+}$) occupy the positions 4*b*, the *f* – ions (R$^{3+}$) occupy the position 4*c*, oxygen ions occupy the positions 4*c* and 8*d* (in Wyckoff notation).

Magnetic moments of the *d* – ions determined by the vectors $M_i$, (*i*=1-4) constitute 4 transition metal magnetic sublattices and the magnetic moments of the *f* – ions determined by vectors $m_i$ (*i*=1-4) constitute 4 rare earth magnetic sublattices. The combinations between magnetic moments of the *d* – and the *f* – sublattices determine magnetic modes of the '*d*' ions

$$\mathbf{F} = \mathbf{M}_1 + \mathbf{M}_2 + \mathbf{M}_3 + \mathbf{M}_4, \quad \mathbf{A} = \mathbf{M}_1 - \mathbf{M}_2 - \mathbf{M}_3 + \mathbf{M}_4$$
$$\mathbf{G} = \mathbf{M}_1 - \mathbf{M}_2 + \mathbf{M}_3 - \mathbf{M}_4, \quad \mathbf{C} = \mathbf{M}_1 + \mathbf{M}_2 - \mathbf{M}_3 - \mathbf{M}_4$$
(7)

and magnetic modes of the '*f*' - ions

$$\mathbf{f} = \mathbf{m}_1 + \mathbf{m}_2 + \mathbf{m}_3 + \mathbf{m}_4, \quad \mathbf{a} = \mathbf{m}_1 - \mathbf{m}_2 - \mathbf{m}_3 + \mathbf{m}_4,$$
$$\mathbf{g} = \mathbf{m}_1 - \mathbf{m}_2 + \mathbf{m}_3 - \mathbf{m}_4, \quad \mathbf{c} = \mathbf{m}_1 + \mathbf{m}_2 - \mathbf{m}_3 - \mathbf{m}_4$$
(8)

Neutron diffraction measurements showed that RCrO$_3$ exhibits one of three *G* – type AFM configurations with weak magnetic component $\Gamma_1(A_x, G_y, C_z)$, $\Gamma_2$ ($F_x, C_y, G_z$), $\Gamma_4$ ($G_x A_y F_z$).

In contrast to BiFeO$_3$, the ferroelectric properties of RCrO$_3$ are rather weak, and the Curie ordering temperatures are low (T$_C$~130 – 250 K) [55,56]. However, as recent experiments have shown, during the polishing of samples in an electric field, an electric polarization of the order of 0.06 µC/cm² is induced; moreover, the electric field affects the spin – reorientation phase transitions, which indicates the presence of magnetoelectric coupling in the material [57,58].

As expected, the magnetoelectric behavior is owed to the crystallographic distortions, which are quite different from those in BiFeO$_3$. In the case of RCrO$_3$ the principal distortions are related to the antirotation of the CrO$_6$ octahedrons around [110] direction and the displacements of R ions from A - positions in perovskite parent phase. Octahedron rotation is described by the axial vectors $\omega_i$ (*i*=1 ÷ 4) linked to Cr ions, and, as was shown, in Ref. [10] they contribute to the axial order parameter

$$\Omega_b = \omega_1 - \omega_2 + \omega_3 - \omega_4$$
(9)

Polar order parameters related to the displacements of oxygen and rare earth ions are also allowed. To compose the polar order parameters, we consider the electric dipole moments $\mathbf{d}_i = q\mathbf{r}_i$ emerging in the vicinity of the Cr ions, where $\mathbf{r}_q = \sum_i q_i \mathbf{r}_{qi} / \sum_i q_i$ is the position of the electric dipole charge center, $q_i$ are the signed magnitudes of the charges, $r_{qi}$ are the radius vectors of the charges in the local reference frame. For the perovskite – like compounds

$$\mathbf{r}_q = \frac{\left(+\frac{3}{8}e\right) \cdot \sum_{i=1}^{8} \mathbf{r}_R + \left(-\frac{2}{2}e\right) \cdot \sum_{i=1}^{6} \mathbf{r}_O}{\left|8 \cdot \left(+\frac{3}{8}e\right) + 6 \cdot \left(-\frac{2}{2}e\right)\right|} \quad \mathbf{r} = (x, y, z)$$
(10)

where *e* is the elementary charge, $r_R$ are the radius vectors of the rare earth ions, $r_O$ are the radius vectors of the oxygen ions measured from Cr$^{3+}$ ion.

As in the case of magnetic moments (6), (7), electric dipoles $\mathbf{d}_i = q\mathbf{r}_i$ constitute 4 ferroelectric sublattices, the combinations between dipole moments determine ferroelectric modes reads

$$P = d_1 + d_2 + d_3 + d_4$$
$$Q_2 = d_1 - d_2 - d_3 + d_4$$
$$Q_3 = d_1 - d_2 + d_3 - d_4$$
$$D = d_1 + d_2 - d_3 - d_4$$

(11)

The arrangement of the dipole moments in RCrO$_3$ is shown in Fig.1; as can be seen, they form an antiferroelectric structure ordered according to the $D$ – mode. Direct calculation of the dipole moments shows that the value of $D$ is maximum, and the values of $P$, $Q_{2,3}$ are negligible.

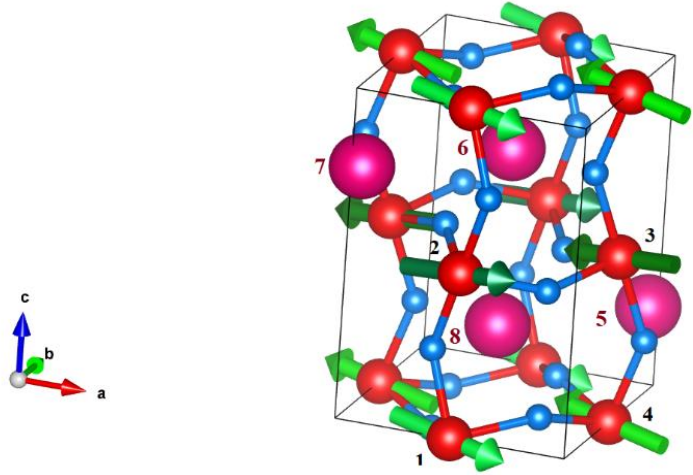

**Figure 1.** Electric dipole moments arrangement in RCrO$_3$ unit cell. Green arrows denote the orientation of electric dipole moments in the vicinity of Cr$^{3+}$ ions ordered by antiferroelectric $D$ mode.

The transformation properties of the order parameters introduced above can be found using the irreducible representations of the space symmetry group of the RCrO$_3$ compounds given in Table 2.

**Table 2.** Irreducible representations of the *Pnma* symmetry group.

| $\Gamma_i$ | $\bar{1}$ | $2_x$ | $2_y$ | $2_z$ | Magnetic OPs, magnetic field | | | Structural OPs, electric field | |
|---|---|---|---|---|---|---|---|---|---|
| | | | | | | 4b | 4c | | |
| $\Gamma_1$ | 1 | 1 | 1 | 1 | $A_z, G_x, C_y$ | $c_y$ | | $\Omega_{bx}$ | |
| $\Gamma_2$ | 1 | 1 | −1 | −1 | $F_z, G_y, C_x, H_z$ | $f_z, c_x$ | | $\Omega_{by}$ | |
| $\Gamma_3$ | 1 | −1 | 1 | −1 | $F_x, A_y, C_z, H_x$ | $f_x, c_z$ | | | |
| $\Gamma_4$ | 1 | −1 | −1 | 1 | $F_y, A_x, G_z, H_y$ | $f_y$ | | $\Omega_{bz}$ | |
| $\Gamma_5$ | −1 | 1 | 1 | 1 | | $g_z, a_x$ | | $Q_{2z}, Q_{3x}, \underline{D_z}$ | |
| $\Gamma_6$ | −1 | 1 | −1 | −1 | | $a_y$ | | $P_z, E_z, Q_{3y}$ | |
| $\Gamma_7$ | −1 | −1 | 1 | −1 | | $a_z, g_x$ | | $P_y, Q_{2x}, E_y, \underline{D_x}$ | |
| $\Gamma_8$ | −1 | −1 | −1 | 1 | | $g_y$ | | $P_x, Q_{2y}, E_x, \underline{D_y}$ | |

Knowledge of their transformation properties makes it possible to understand how crystallographic distortions manifest themselves in magnetic and ferroelectric orderings. As in the previous subsection, classification of the components of order parameters according to the irreducible representations of the *Pnma* symmetry group allows us to determine exchange – coupled magnetic structures, the possible electrostatically – coupled dipole structures, to compose invariant combination of order parameters contributing to thermodynamic potential and calculate magnetoelectric coupling coefficient.

$P_x=(\alpha_{xyx}a_y+ \alpha_{xzyx}g_zG_yG_x + \alpha_{xxyx}a_xG_y G_x )H_x+ (\alpha_{xxy}a_x+\alpha_{xxyy}f_xG_y G_x + \alpha_{xzyy}c_zG_y G_x + \alpha_{xzy}g_z)H_y +…$

$F_x=(\alpha^*_{xyx}a_y+ \alpha^*_{xzyx}g_zG_y G_x + \alpha^*_{xxyx}a_xG_y G_x )E_x+ (\alpha^*_{xxy}a_x+\alpha^*_{xxyy}f_xG_y G_x + \alpha^*_{xzyy}c_zG_yG_x + \alpha^*_{xzy}g_z)E_y +…$  (12)

Magnetoelectric coefficient in terms of the components of G and $g$ vectors is written as follows

$$\alpha_{ij}^{RCrO_3} = \begin{vmatrix} a_1G_x(g_zG_y - g_yG_z) & a_2g_z & a_3g_x \\ a_2g_z & a_1G_x(g_zG_y - g_yG_z) & a_4g_y \\ a_4g_y & a_3g_x & a_2g_z \end{vmatrix}$$

(13)

At the end of this section, we draw the reader's attention to the fact that the presented analysis applies to rare-earth orthoferrites $RFeO_3$. $RFeO_3$ compounds crystallize in the same space symmetry group Pnma, and several of these crystals exhibit non-linear magnetoelectric responses [23,25,59,59,60]. The magnetoelectric tensor (12) can be also used to describe their magnetoelectric properties.

### 2.3. Ruddlesden – Popper structures

Ruddlesden-Popper (RP) structures are attracting considerable attention due to the recently discovered high-temperature ferroelectric properties and the prospects for their potential implementation as multiferroics at room temperature [11]. The presence of ferroelectric properties, the so - called hybrid improper ferroelectricity (HIF), was theoretically predicted in double – layered perovskite compounds $Ca_3Mn_2O_7$ and $Ca_3Ti_2O_7$, and then detected experimentally in $Ca_{3-x}Sr_xTi_2O_7$ single crystals [20], and in $Ca_3(Ti_{1-x}Mn_x)_2O_7$ ceramics [21].

In this section, we plan to focus on the RP manganite phases $Ca_3(Ti_{1-x}Mn_x)_2O_7$, since they contain magnetic Mn ions and, as shown in Ref. [21], are room – temperature HIF materials. In the limiting case $x=1$, $Ca_3(Ti_{1-x}Mn_x)_2O_7$ transforms into the RP phase $Ca_3Mn_2O_7$, which was thoroughly explored due to the colossal magnetoresistance effect found in the $(CaO)-(CaMnO_3)_n$ ($n =1, 2, 3, \infty$) RP structures in the 2000s [16–19]. The structures with $n = 1$ (2D - structure) and $n = \infty$ (perovskite) are considered as the end members of the RP series, so the double-layered RP structures receive more attention.

The temperature of AFM ordering $T_N = 115$ K in $Ca_3Mn_2O_7$ was first experimentally found in Ref. [16], there it was assumed that there is a G – type AFM order. Neutron diffraction study [19] revealed the dominant contribution of G – type or C – type AFM phase with an AFM spin arrangement within the bi – layer plane. In that work, the possibility of the existence of a weak ferromagnetic (WFM) state in the RP phase $Ca_3Mn_2O_7$ was also assumed. Further study on $Ca_3Mn_2O_7$ showed the transition to the AFM state at the temperature $T_N = 134$ K and signature of WFM below 100 K [61,62].

The crystal structure, structural phase transitions, and polar phases of $Ca_3Ti_2O_7$ [63] and $Ca_3Mn_2O_7$ [19,64] with double-layered RP structures, studied since 1998 [16], have now been proven [21]. Above room temperature (RT) the crystal structure of $Ca_3Mn_2O_7$ is described by the tetragonal space group *I4/mmm*, and at RT, by the space group *Cmc*$2_1$ [64]. The transition from *I4/mmm* to *Cmc*$2_1$ phase should occur through an intermediate phase, which can be either *Cmcm* or *Cmca*, as follows from the first-principles calculations [46]. The transition into ferroelectric phase is observed close to $T_C = 280$ K, at which the transition into *Cmc*$2_1$ phase occurs. Since experimental measurements indicate that ferroelectric ordering in $Ca_3Mn_2O_7$ occurs up to RT, the existence of an intermediate ferroelectric state is expected here.

Structural transformations can arise due to the substitution of magnetic ions at the A – positions ($Ca_{3-x}La_xMn_2O_7$) or ferroelectric ions in the B – positions ($Ca_3(Ti_{1-x}Mn_x)_2O_7$). The tetragonal-to-orthorhombic transition presumable through the phases *I4/mmm* → *Fmmm* → *Cmcm* → *Cmc*$2_1$, occurring in the 200–300 °C range, was observed during the study of $Ca_{3-x}La_xMn_2O_7$ structures [19,65]. Though the existence of *I4/mmm* and *Cmc*$2_1$ phases is confirmed, the presence of 'intermediate' phases is still under discussion.

So, in further consideration, we can take the tetragonal *I4/mmm* structure ($Ca_3Mn_2O_7$) as the parent phase and apply the symmetry analysis, employed in the previous sections, to explore the magnetoelectric properties of the RP structures. The atomic coordinates in *I4/mmm* phase obtained from the synchrotron x-ray diffraction study [19] are given in **Table 3**.

**Table 3.** Positional and thermal parameters of atoms and Mn–O interatomic distances (Å) in I4/mmm phase $Ca_3Mn_2O_7$, a=3.69185 (5), c=19.6254(5) [19].

| atom | Wycoff position | x | y | z |
|---|---|---|---|---|
| Ca1 | 2b | 0 | ½ | 0.7(2) |
| Ca2 | 4e | 0 | 0.314 | 0.5(2) |
| Mn | 4e | 0 | 0.097 | 0.5(1) |
| O1 | 2a | 0 | 0 | 2.2(2) |
| O2 | 4e | 0 | 0.194(1) | =B(O1) |
| O3 | 8g | ½ | 0.102(9) | =B(O1) |

It is seen in **Table 3** that the Mn ions occupy the 4e positions (in the Wyckoff notation), which indicates the presence of 4 magnetic sublattices $\mu_i$ ($i=1\div4$) with magnetic moments of the same magnitude. Possible combinations of the vectors $\mu_i$ give us the basic magnetic vectors

$$F = \mu_1 + \mu_2 + \mu_3 + \mu_4$$
$$A = \mu_1 - \mu_2 - \mu_3 + \mu_4$$
$$G = \mu_1 - \mu_2 + \mu_3 - \mu_4$$
$$C = \mu_1 + \mu_2 - \mu_3 - \mu_4 \quad (14)$$

The unit cell of $Ca_3Mn_2O_7$ shown in Figure 2, contains 2 formula units $Ca_3Mn_2O_7$. The positions of the symmetry elements $I, C_2^z, C_4^z$ in the unit cell are depicted in Fig. 2b in accordance with Figure 2a. Note that the local symmetry elements of the 4e{4mm} positions, which are the axis $4_z$ (principle crystal axis) and 4 planes of symmetry *m* passing through this axis; remain even for magnetic ions, occupying the 4e positions.

In accordance with Figure 3, it is fairly easy to compose permutation transformations of the ions at the 4e positions (**Table 4**) and permutation transformations of basis vectors (**Table 5**).

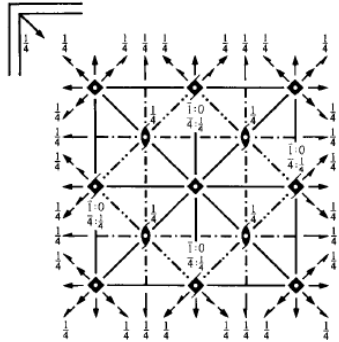
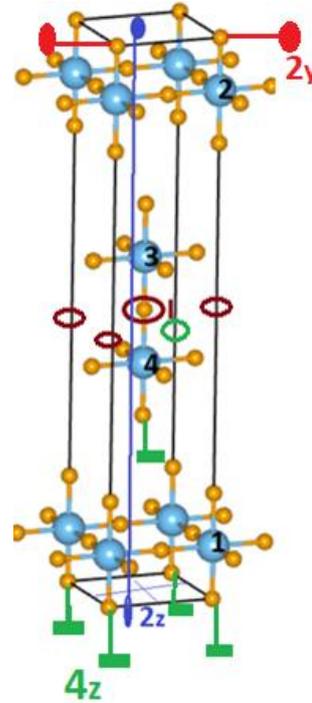

*a*         *b*

**Figure 2.** a) *I4/mmm* space symmetry group, b) Ca3Mn2O7 unit cell.

**Table 4.** Permutation transformations ions in positions 4e {4mm}.

| $G_F$ | 1 | 2 | 3 | 4 |
|---|---|---|---|---|
| $\bar{1}$ | 2 | 1 | 4 | 3 |
| $4_z$ | 1 | 2 | 3 | 4 |
| $2_z$ | 3 | 4 | 1 | 2 |
| $2_y$ | 2 | 1 | 4 | 3 |

**Table 5**. Permutation transformations of basis vectors.

| $G_F$ | F | A | G | C |
|---|---|---|---|---|
| $\bar{1}$ | F | −A | −G | C |
| $4_z$ | F | A | G | C |
| $2_z$ | F | −A | G | −C |
| $2_y$ | F | −A | −G | C |

Using **Table 5**, we obtain a cipher (Turov indices) for AFM vectors, which explains how the symmetry elements transform magnetic sublattices into each other

$$A \quad 4_z(+)2_z^1(-)2_y(-)\bar{1}(-)$$

$$G \quad 4_z(+)2_z^1(+)2_y(-)\bar{1}(-)$$

$$C \quad 4_z(+)2_z^1(-)2_y(+)\bar{1}(+)$$

In consistence with the data of experimental studies [16,19], we assume the G – type of AFM ordering. In this case, the values of AFM vectors *A* and *C* are taken to be sufficiently small and we can restrict our consideration with two magnetic order parameters instead of 4. So, we leave the ferromagnetic vector *M=F* and the antiferromagnetic vector *L=G* as the basic magnetic parameters. Then we classify them together with polarization vector according to the irreducible representation of the space symmetry group $D_{4h}^{17}$ (**Table 6**). The first line of Table 6 contains generators of $D_{4h}^{17}$ group $G = \{E, I, C_2^y, C_2^z, C_4^{z+}\}$, the last column contains basic magnetic and ferroelectric functions.

The decomposition of the basic functions into irreducible representations of the symmetry group *I*4/*mmm* allows one to get information of the properties of a system. It is seen that the ferromagnetic and antiferromagnetic vectors transform according to the different IRs, an external factors, such as electric field or the strain will allow to transform *F* and *L* into each other, the combinations $F_z(P_xG_x + P_yG_y), G_z(F_xP_x + F_yP_y), P_z(F_xG_x + F_yP_zG_y), F_zP_zG_z$, transforming on Γ1, are invariants and give impact into magnetoelectric energy

$$\Phi_{me} = \gamma_2 F_z\left(P_xG_x + P_yG_y\right) + \gamma_3 G_z\left(F_xP_x + F_yP_y\right) + \gamma_4 P_z\left(F_xG_x + F_yG_y\right) + \gamma_5 F_zP_zG_z, \qquad (15)$$

where
$$\gamma_2 = \gamma_{zxx} = \gamma_{zyy}, \gamma_3 = \gamma_{xxz} = \gamma_{yyz}, \gamma_4 = \gamma_{xzx} = \gamma_{yzy}, \gamma_5 = \gamma_{zzz}$$

This finding allows us to conclude that the linear magnetoelectric effect is allowed by symmetry of the RP structures.

**Table 6.** Irreducible representations of the *I*4/*mmm* symmetry group and basic functions

| IR | E | $2C_{4z}$ | $C_2^z$ | $2C_2^y$ | I | basis vectors |
|---|---|---|---|---|---|---|
| | | | | | | **F, G, P** |
| Γ1 | 1 | 1 | 1 | 1 | 1 | |
| Γ2 | 1 | 1 | 1 | −1 | 1 | $F_z$ |
| Γ3 | 1 | −1 | 1 | −1 | 1 | |
| Γ4 | 1 | −1 | 1 | 1 | 1 | |
| Γ5 | $\begin{pmatrix}1&0\\0&1\end{pmatrix}$ | $\begin{pmatrix}0&1\\-1&0\end{pmatrix}$ | $\begin{pmatrix}-1&0\\0&-1\end{pmatrix}$ | $\begin{pmatrix}-1&0\\0&1\end{pmatrix}$ | $\begin{pmatrix}1&0\\0&1\end{pmatrix}$ | $\begin{pmatrix}F_x\\F_y\end{pmatrix}$ |
| Γ6 | 1 | 1 | 1 | 1 | −1 | $G_z$ |
| Γ7 | 1 | 1 | 1 | −1 | −1 | $P_z$ |
| Γ8 | 1 | −1 | 1 | −1 | −1 | |
| Γ9 | 1 | −1 | 1 | 1 | −1 | |
| Γ10 | $\begin{pmatrix}1&0\\0&1\end{pmatrix}$ | $\begin{pmatrix}0&1\\-1&0\end{pmatrix}$ | $\begin{pmatrix}-1&0\\0&-1\end{pmatrix}$ | $\begin{pmatrix}-1&0\\0&1\end{pmatrix}$ | $\begin{pmatrix}-1&0\\0&-1\end{pmatrix}$ | $\begin{pmatrix}P_x\\P_y\end{pmatrix}$ |
| Γ10' | $\begin{pmatrix}1&0\\0&1\end{pmatrix}$ | $\begin{pmatrix}0&1\\-1&0\end{pmatrix}$ | $\begin{pmatrix}-1&0\\0&-1\end{pmatrix}$ | $\begin{pmatrix}1&0\\0&-1\end{pmatrix}$ | $\begin{pmatrix}-1&0\\0&-1\end{pmatrix}$ | $\begin{pmatrix}G_x\\G_y\end{pmatrix}$ |

$$\alpha_{ij}^{R-P} = \begin{vmatrix} \gamma_3 G_z & 0 & \gamma_2 G_x \\ 0 & \gamma_3 G_z & \gamma_2 G_y \\ \gamma_4 G_x & \gamma_4 G_y & \gamma_5 G_z \end{vmatrix}$$

(16)

### 3. Results and conclusion

To summarize, we apply the symmetry analysis to the magnetic and magnetoelectric properties of multiferroics (MFs) with a perovskite structure. One of the reasons for choosing a perovskite – based MFs as the object of this study is related to the structural instability of the initial perovskite phase, which can lead to a variety of structures with different symmetry. This allows us to demonstrate how crystallographic distortions of various types, even if they are small enough, significantly alter the magnetic and ferroelectric orderings and their couplings. As the typical examples, we consider i) the family of high – temperature multiferroics $Bi_x R_{1-x}FeO_3$ recognizable by their magnetoelectric properties; ii) the rare – earth orthochromites $RCrO_3$, promising candidates for MFs, compounds with a well - known magnetic structures, iii) the RP structures containing magnetic cations, such as $Ca_3(Ti_{1-x}Mn_x)_2O_7$, novel high – temperature MFs, the ferroelectric and magnetic properties of which are still being researched. These structures crystallize into trigonal, rhombic, and tetragonal syngonies, which allows us to trace the relationship between magnetoelectric properties and symmetry of a structure and consider impacts given by crystallographic distortions.

The direct contribution of the distortion into electric polarization and weak ferromagnetism was considered, using $BiFeO_3$ as an example. We also analyzed the influence of structural transformation the $Bi_xR_{1-x}FeO_3$ family on their magnetoelectric properties. Calculations show that the linear magnetoelectric effect, suppressed by the spin modulated structure in pure $BiFeO_3$, becomes allowed due to the symmetry of the new phase in $Bi_xR_{1-x}FeO_3$, where it is attributed mainly to R ions.

In the case of the rare – earth orthochromites $RCrO_3$ it has been shown that the ferroelectric and magnetoelectric properties are due to crystallographic distortions. The displacements of oxygen ions from their positions in the initial perovskite phase results in the emergence of electric dipole moments in the vicinity of the Cr ions, which are coupled with the magnetic moments of the Cr ions. The distortive, ferroelectric and magnetic order parameters have been classified according to irreducible representations of the Pnma symmetry group of RCrO$_3$, which allows to compose invariant combinations between these parameters.

For the first time, the similar symmetry consideration was implemented to the analysis of the Ruddlesden – Popper structures. Taking into account, the local symmetry of magnetic ions in the RP unit cell, we introduced the relevant magnetic order parameters and classified them according to the irreducible representation of the I4/mmm symmetry group, which describes the tetragonal symmetry of the RP structures. Analysis exemplified on the Ca$_3$(Ti$_{1-x}$Mn$_x$)$_2$O$_7$ compounds demonstrates the possibility of realizing MEE in the RP phases containing magnetically active cations and allows estimate the magnetoelectric contribution to the thermodynamic potential.

In conclusion, this research allowed us to compare the magnetoelectric effects for different crystal systems of perovskites and thus design a more meaningful organization of the desired experiments.

**Author Contributions:** Conceptualization, A.K.Z, Z.V.G, X.M.C.; methodology, K.A.Z; formal analysis, Z.V.G; writing—Z.V.G., A.K.Z., K.A.Z., X.M.C.

**Funding:** Z.V.G. acknowledges the support the State assignment for the implementation of scientific research by laboratories (Order MN-8/1356 of 09/20/2021), Russian Foundation for Basic Research under grant No. 19-52-80024, A.K.Z. and K.A.Z. acknowledge the f, Russian Foundation for Basic Research under grant No. 19-52-80024, X.M.C. acknowledges the National Natural Science Foundation of China under Grant No. 51961145105.

**Conflicts of Interest:** The authors declare no conflict of interest.

**References**

(1) Manipatruni, S.; Nikonov, D. E.; Lin, C.-C.; Gosavi, T. A.; Liu, H.; Prasad, B.; Huang, Y.-L.; Bonturim, E.; Ramesh, R.; Young, I. A. Scalable Energy-Efficient Magnetoelectric Spin–Orbit Logic. *Nature* **2019**, *565* (7737), 35–42. https://doi.org/10.1038/s41586-018-0770-2.

(2) Dubrovin, R. M.; Alyabyeva, L. N.; Siverin, N. V.; Gorshunov, B. P.; Novikova, N. N.; Boldyrev, K. N.; Pisarev, R. V. Incipient Multiferroicity in $Pnma$ Fluoroperovskite ${\mathrm{NaMnF}}_{3}$. *Phys. Rev. B* **2020**, *101* (18), 180403. https://doi.org/10.1103/PhysRevB.101.180403.

(3) Liu, H.; Yang, X. A Brief Review on Perovskite Multiferroics. *Ferroelectrics* **2017**, *507* (1), 69–85. https://doi.org/10.1080/00150193.2017.1283171.

(4) Ke, X.; Birol, T.; Misra, R.; Lee, J.-H.; Kirby, B. J.; Schlom, D. G.; Fennie, C. J.; Freeland, J. W. Structural Control of Magnetic Anisotropy in a Strain-Driven Multiferroic EuTiO${}_{3}$ Thin Film. *Phys. Rev. B* **2013**, *88* (9), 094434. https://doi.org/10.1103/PhysRevB.88.094434.

(5) Catalan, G.; Scott, J. F. Physics and Applications of Bismuth Ferrite. *Adv. Mater.* **2009**, *21* (24), 2463–2485. https://doi.org/10.1002/adma.200802849.

(6) Lin, P.-T.; Li, X.; Zhang, L.; Yin, J.-H.; Cheng, X.-W.; Wang, Z.-H.; Wu, Y.-C.; Wu, G.-H. La-Doped BiFeO 3 : Synthesis and Multiferroic Property Study. *Chin. Phys. B* **2014**, *23* (4), 047701. https://doi.org/10.1088/1674-1056/23/4/047701.

(7) Savosta, M. M.; Novák, P.; Maryško, M.; Jirák, Z.; Hejtmánek, J.; Englich, J.; Kohout, J.; Martin, C.; Raveau, B. Coexistence of Antiferromagnetism and Ferromagnetism in ${\mathrm{Ca}}_{1\ensuremath{-}x}{\mathrm{Pr}}_{x}{\mathrm{MnO}}_{3}$ (X<~0.1)$ Manganites. *Phys. Rev. B* **2000**, *62* (14), 9532–9537. https://doi.org/10.1103/PhysRevB.62.9532.

(8) Zvezdin, A. K.; Mukhin, A. A. Magnetoelectric Interactions and Phase Transitions in a New Class of Multiferroics with Improper Electric Polarization. *JETP Lett.* **2008**, *88* (8), 505–510. https://doi.org/10.1134/S0021364008200083.

(9) Shikin, A. M.; Estyunin, D. A.; Zaitsev, N. L.; Glazkova, D.; Klimovskikh, I. I.; Filnov, S. O.; Rybkin, A. G.; Schwier, E. F.; Kumar, S.; Kimura, A.; Mamedov, N.; Aliev, Z.; Babanly, M. B.; Kokh, K.; Tereshchenko, O. E.; Otrokov, M. M.; Chulkov, E. V.; Zvezdin, K. A.; Zvezdin, A. K. Sample-Dependent Dirac-Point Gap in ${\mathrm{MnBi}}_{2}{\mathrm{Te}}_{4}$ and Its Response to Applied Surface Charge: A Combined Photoemission and Ab Initio Study. *Phys. Rev. B* **2021**, *104* (11), 115168. https://doi.org/10.1103/PhysRevB.104.115168.

(10) Zvezdin, A. K.; Gareeva, Z. V.; Chen, X. M. Multiferroic Order Parameters in Rhombic Antiferromagnets RCrO3. *J. Phys. Condens. Matter* **2021**, *33* (38), 385801. https://doi.org/10.1088/1361-648X/ac0dd6.

(11) Zhang, B. H.; Liu, X. Q.; Chen, X. M. Review of Experimental Progress of Hybrid Improper Ferroelectricity in Layered Perovskite Oxides. *J. Phys. Appl. Phys.* **2021**, *55* (11), 113001. https://doi.org/10.1088/1361-6463/ac3284.

(12) Evans, H. A.; Mao, L.; Seshadri, R.; Cheetham, A. K. Layered Double Perovskites. *Annu. Rev. Mater. Res.* **2021**, *51* (1), 351–380. https://doi.org/10.1146/annurev-matsci-092320-102133.

(13) Ruddlesden, S. N.; Popper, P. The Compound Sr3Ti2O7 and Its Structure. *Acta Crystallogr.* **1958**, *11* (1), 54–55. https://doi.org/10.1107/S0365110X58000128.

(14) Ding, P.; Li, W.; Zhao, H.; Wu, C.; Zhao, L.; Dong, B.; Wang, S. Review on Ruddlesden–Popper Perovskites as Cathode for Solid Oxide Fuel Cells. *J. Phys. Mater.* **2021**, *4* (2), 022002. https://doi.org/10.1088/2515-7639/abe392.

(15) Saw, A. K.; Gupta, S.; Dayal, V. Structural, Magneto Transport and Magnetic Properties of Ruddlesden–Popper La2-2xSr1+2xMn2O7 (0.42≤x≤0.52) Layered Manganites. *AIP Adv.* **2021**, *11* (2), 025331. https://doi.org/10.1063/9.0000109.

(16) Battle, P. D.; Rosseinsky*, M. J. Synthesis, Structure, and Magnetic Properties of N=2 Ruddlesden–Popper Manganates. *Curr. Opin. Solid State Mater. Sci.* **1999**, *4* (2), 163–170. https://doi.org/10.1016/S1359-0286(99)00012-1.

(17) Fawcett, I. D.; Sunstrom; Greenblatt, M.; Croft, M.; Ramanujachary, K. V. Structure, Magnetism, and Properties of Ruddlesden−Popper Calcium Manganates Prepared from Citrate Gels. *Chem. Mater.* **1998**, *10* (11), 3643–3651. https://doi.org/10.1021/cm980380b.

(18) Harris, A. B. Symmetry Analysis for the Ruddlesden-Popper Systems $Ca_{3}Mn_{2}O_{7}$ and $Ca_{3}Ti_{2}O_{7}$. *Phys. Rev. B* **2011**, *84* (6), 064116. https://doi.org/10.1103/PhysRevB.84.064116.

(19) Lobanov, M. V.; Greenblatt, M.; Caspi, E. ad N.; Jorgensen, J. D.; Sheptyakov, D. V.; Toby, B. H.; Botez, C. E.; Stephens, P. W. Crystal and Magnetic Structure of the Ca3Mn2O7 Ruddlesden–Popper Phase: Neutron and Synchrotron x-Ray Diffraction Study. *J. Phys. Condens. Matter* **2004**, *16* (29), 5339–5348. https://doi.org/10.1088/0953-8984/16/29/023.

(20) Oh, Y. S.; Luo, X.; Huang, F.-T.; Wang, Y.; Cheong, S.-W. Experimental Demonstration of Hybrid Improper Ferroelectricity and the Presence of Abundant Charged Walls in (Ca,Sr)3Ti2O7 Crystals. *Nat. Mater.* **2015**, *14* (4), 407–413. https://doi.org/10.1038/nmat4168.

(21) Liu, X. Q.; Wu, J. W.; Shi, X. X.; Zhao, H. J.; Zhou, H. Y.; Qiu, R. H.; Zhang, W. Q.; Chen, X. M. Hybrid Improper Ferroelectricity in Ruddlesden-Popper Ca3(Ti,Mn)2O7 Ceramics. *Appl. Phys. Lett.* **2015**, *106* (20), 202903. https://doi.org/10.1063/1.4921624.

(22) Glazer, A. M. The Classification of Tilted Octahedra in Perovskites. *Acta Crystallogr. B* **1972**, *28* (11), 3384–3392. https://doi.org/10.1107/S0567740872007976.

(23) Senn, M. S.; Bristowe, N. C. A Group-Theoretical Approach to Enumerating Magnetoelectric and Multiferroic Couplings in Perovskites. *Acta Crystallogr. Sect. Found. Adv.* **2018**, *74* (4), 308–321. https://doi.org/10.1107/S2053273318007441.

(24) Mulder, A. T.; Benedek, N. A.; Rondinelli, J. M.; Fennie, C. J. Turning ABO3 Antiferroelectrics into Ferroelectrics: Design Rules for Practical Rotation-Driven Ferroelectricity in Double Perovskites and A3B2O7 Ruddlesden-Popper Compounds. *Adv. Funct. Mater.* **2013**, *23* (38), 4810–4820. https://doi.org/10.1002/adfm.201300210.

(25) Bousquet, E.; Cano, A. Non-Collinear Magnetism in Multiferroic Perovskites. *J. Phys. Condens. Matter* **2016**, *28* (12), 123001. https://doi.org/10.1088/0953-8984/28/12/123001.

(26) Narayanan, N.; Graham, P. J.; Rovillain, P.; O'Brien, J.; Bertinshaw, J.; Yick, S.; Hester, J.; Maljuk, A.; Souptel, D.; Büchner, B.; Argyriou, D.; Ulrich, C. Reduced Crystal Symmetry as Origin of the Ferroelectric Polarization within the Incommensurate Magnetic Phase of TbMn2O5. *ArXiv210905164 Cond-Mat* **2021**.

(27) Perez-Mato, J. M.; Ribeiro, J. L.; Petricek, V.; Aroyo, M. I. Magnetic Superspace Groups and Symmetry Constraints in Incommensurate Magnetic Phases. *J. Phys. Condens. Matter* **2012**, *24* (16), 163201. https://doi.org/10.1088/0953-8984/24/16/163201.

(28) Hatch, D. M.; Stokes, H. T. INVARIANTS: Program for Obtaining a List of Invariant Polynomials of the Order-Parameter Components Associated with Irreducible Representations of a Space Group. *J. Appl. Crystallogr.* **2003**, *36* (3), 951–952. https://doi.org/10.1107/S0021889803005946.

(29) (International Tables for Crystallography) Introduction to the properties of tensors https://onlinelibrary.wiley.com/iucr/itc/Da/ch1o1v0001/ (accessed 2021 -12 -27).

(30) Gabbasova, Z. V.; Kuz'min, M. D.; Zvezdin, A. K.; Dubenko, I. S.; Murashov, V. A.; Rakov, D. N.; Krynetsky, I. B. Bi1−xRxFeO3 (R=rare Earth): A Family of Novel Magnetoelectrics. *Phys. Lett. A* **1991**, *158* (9), 491–498. https://doi.org/10.1016/0375-9601(91)90467-M.

(31) Palkar, V. R.; John, J.; Pinto, R. Observation of Saturated Polarization and Dielectric Anomaly in Magnetoelectric BiFeO3 Thin Films. *Appl. Phys. Lett.* **2002**, *80* (9), 1628–1630. https://doi.org/10.1063/1.1458695.

(32) Lebeugle, D.; Colson, D.; Forget, A.; Viret, M.; Bonville, P.; Marucco, J. F.; Fusil, S. Room-Temperature Coexistence of Large Electric Polarization and Magnetic Order in $\mathrm{Bi}\mathrm{Fe}{\mathrm{O}}_{3}$ Single Crystals. *Phys. Rev. B* **2007**, *76* (2), 024116. https://doi.org/10.1103/PhysRevB.76.024116.

(33) Lebeugle, D.; Colson, D.; Forget, A.; Viret, M. Very Large Spontaneous Electric Polarization in BiFeO3 Single Crystals at Room Temperature and Its Evolution under Cycling Fields. *Appl. Phys. Lett.* **2007**, *91* (2), 022907. https://doi.org/10.1063/1.2753390.

(34) Teague, J. R.; Gerson, R.; James, W. J. Dielectric Hysteresis in Single Crystal BiFeO3. *Solid State Commun.* **1970**, *8* (13), 1073–1074. https://doi.org/10.1016/0038-1098(70)90262-0.

(35) Wang, Y. P.; Yuan, G. L.; Chen, X. Y.; Liu, J.-M.; Liu, Z. G. Electrical and Magnetic Properties of Single-Phased and Highly Resistive Ferroelectromagnet BiFeO3ceramic. *J. Phys. Appl. Phys.* **2006**, *39* (10), 2019–2023. https://doi.org/10.1088/0022-3727/39/10/006.

(36) Li, J.; Wang, J.; Wuttig, M.; Ramesh, R.; Wang, N.; Ruette, B.; Pyatakov, A. P.; Zvezdin, A. K.; Viehland, D. Dramatically Enhanced Polarization in (001), (101), and (111) BiFeO3 Thin Films Due to Epitiaxial-Induced Transitions. *Appl. Phys. Lett.* **2004**, *84* (25), 5261–5263. https://doi.org/10.1063/1.1764944.

(37) Dixit, H.; Beekman, C.; Schlepütz, C. M.; Siemons, W.; Yang, Y.; Senabulya, N.; Clarke, R.; Chi, M.; Christen, H. M.; Cooper, V. R. Understanding Strain-Induced Phase Transformations in BiFeO3 Thin Films. *Adv. Sci.* **2015**, *2* (8), 1500041. https://doi.org/10.1002/advs.201500041.

(38) Sando, D.; Agbelele, A.; Rahmedov, D.; Liu, J.; Rovillain, P.; Toulouse, C.; Infante, I. C.; Pyatakov, A. P.; Fusil, S.; Jacquet, E.; Carrétéro, C.; Deranlot, C.; Lisenkov, S.; Wang, D.; Le Breton, J.-M.; Cazayous, M.; Sacuto, A.; Juraszek, J.; Zvezdin, A. K.; Bellaiche, L.; Dkhil, B.; Barthélémy, A.; Bibes, M. Crafting the Magnonic and Spintronic Response of BiFeO3 Films by Epitaxial Strain. *Nat. Mater.* **2013**, *12* (7), 641–646. https://doi.org/10.1038/nmat3629.

(39) Resta, R.; Vanderbilt, D. Theory of Polarization: A Modern Approach. In *Physics of Ferroelectrics: A Modern Perspective*; Topics in Applied Physics; Springer: Berlin, Heidelberg, 2007; pp 31–68. https://doi.org/10.1007/978-3-540-34591-6_2.

(40) Spaldin, N. A. A Beginner's Guide to the Modern Theory of Polarization. *J. Solid State Chem.* **2012**, *195*, 2–10. https://doi.org/10.1016/j.jssc.2012.05.010.

(41) Gareeva, Z.; Diéguez, O.; Íñiguez, J.; Zvezdin, A. K. Complex Domain Walls in ${\mathrm{BiFeO}}_{3}$. *Phys. Rev. B* **2015**, *91* (6), 060404. https://doi.org/10.1103/PhysRevB.91.060404.

(42) Zvezdin, A. K.; Pyatakov, A. P. On the Problem of Coexistence of the Weak Ferromagnetism and the Spin Flexoelectricity in Multiferroic Bismuth Ferrite. *EPL Europhys. Lett.* **2012**, *99* (5), 57003. https://doi.org/10.1209/0295-5075/99/57003.

(43) Fishman, R. S.; Rõõm, T.; de Sousa, R. Normal Modes of a Spin Cycloid or Helix. *Phys. Rev. B* **2019**, *99* (6), 064414. https://doi.org/10.1103/PhysRevB.99.064414.

(44) Sosnowska, I.; Neumaier, T. P.; Steichele, E. Spiral Magnetic Ordering in Bismuth Ferrite. *J. Phys. C Solid State Phys.* **1982**, *15* (23), 4835–4846. https://doi.org/10.1088/0022-3719/15/23/020.

(45) Kadomtseva, A. M.; Zvezdin, A. K.; Popov, Yu. F.; Pyatakov, A. P.; Vorob'ev, G. P. Space-Time Parity Violation and Magnetoelectric Interactions in Antiferromagnets. *J. Exp. Theor. Phys. Lett.* **2004**, *79* (11), 571–581. https://doi.org/10.1134/1.1787107.

(46) Kadomtseva, A. M.; Popov, Yu. F.; Pyatakov, A. P.; Vorob'ev, G. P.; Zvezdin, A. K.; Viehland, D. Phase Transitions in Multiferroic BiFeO3 Crystals, Thin-Layers, and Ceramics: Enduring Potential for a Single Phase, Room-Temperature Magnetoelectric 'Holy Grail.' *Phase Transit.* **2006**, *79* (12), 1019–1042. https://doi.org/10.1080/01411590601067235.

(47) Popov, Yu. F.; Kadomtseva, A. M.; Vorob'ev, G. P.; Zvezdin, A. K. Discovery of the Linear Magnetoelectric Effect in Magnetic Ferroelectric BiFeo3 in a Strong Magnetic Field. *Ferroelectrics* **1994**, *162* (1), 135–140. https://doi.org/10.1080/00150199408245098.
(48) Pyatakov, A. P.; Zvezdin, A. K. Magnetoelectric and Multiferroic Media. *Phys.-Uspekhi* **2012**, *55* (6), 557. https://doi.org/10.3367/UFNe.0182.201206b.0593.
(49) Murashov, V. A.; Rakov, D. N.; Ehkonomov, N. A. Quadratic Magnetoelectric Effect in Monocrystalline (Bi,La)FeO3. *Fiz. Tverd. Tela* **1990**, *32* (7), 2156–2158.
(50) Sosnowska, I.; Przeniosło, R.; Fischer, P.; Murashov, V. A. Neutron Diffraction Studies of the Crystal and Magnetic Structures of BiFeO3 and Bi0.93La0.07FeO3. *J. Magn. Magn. Mater.* **1996**, *160*, 384–385. https://doi.org/10.1016/0304-8853(96)00240-5.
(51) Chen, J.; Liu, L.; Zhu, X. L.; Gareeva, Z. V.; Zvezdin, A. K.; Chen, X. M. The Involvement of Pna21 Phase in the Multiferroic Characteristics of La/Lu Co-Substituted BiFeO3 Ceramics. *Appl. Phys. Lett.* **2021**, *119* (11), 112901. https://doi.org/10.1063/5.0059793.
(52) Belov, K. P.; Zvezdin, A. K.; Kadomtseva, A. M.; Levitin, R. Z. Spin-Reorientation Transitions in Rare-Earth Magnets. *Sov. Phys. Uspekhi* **1976**, *19* (7), 574. https://doi.org/10.1070/PU1976v019n07ABEH005274.
(53) Turov, E. A. Can the Magnetoelectric Effect Coexist with Weak Piezomagnetism and Ferromagnetism? *Phys.-Uspekhi* **1994**, *37* (3), 303. https://doi.org/10.1070/PU1994v037n03ABEH000015.
(54) Izyumov, Yu. A.; Naish, V. E. Symmetry Analysis in Neutron Diffraction Studies of Magnetic Structures: 1. A Phase Transition Concept to Describe Magnetic Structures in Crystals. *J. Magn. Magn. Mater.* **1979**, *12* (3), 239–248. https://doi.org/10.1016/0304-8853(79)90086-6.
(55) Tassel, C.; Goto, Y.; Kuno, Y.; Hester, J.; Green, M.; Kobayashi, Y.; Kageyama, H. Direct Synthesis of Chromium Perovskite Oxyhydride with a High Magnetic-Transition Temperature. *Angew. Chem.* **2014**, *126* (39), 10545–10548. https://doi.org/10.1002/ange.201405453.
(56) Singh, K. D.; Singh, F.; Choudhary, R. J.; Kumar, R. Consequences of $R^{3+}$ Cationic Radii on the Dielectric and Magnetic Behavior of RCrO3 Perovskites. *Appl. Phys. A* **2020**, *126* (3), 148. https://doi.org/10.1007/s00339-020-3324-z.
(57) Sanina, V. A.; Khannanov, B. Kh.; Golovenchits, E. I.; Shcheglov, M. P. Electric Polarization in ErCrO3 Induced by Restricted Polar Domains. *Phys. Solid State* **2019**, *61* (3), 370–378. https://doi.org/10.1134/S1063783419030284.
(58) Rajeswaran, B.; Khomskii, D. I.; Zvezdin, A. K.; Rao, C. N. R.; Sundaresan, A. Field-Induced Polar Order at the N\'eel Temperature of Chromium in Rare-Earth Orthochromites: Interplay of Rare-Earth and Cr Magnetism. *Phys. Rev. B* **2012**, *86* (21), 214409. https://doi.org/10.1103/PhysRevB.86.214409.
(59) Tokunaga, Y.; Furukawa, N.; Sakai, H.; Taguchi, Y.; Arima, T.; Tokura, Y. Composite Domain Walls in a Multiferroic Perovskite Ferrite. *Nat. Mater.* **2009**, *8* (7), 558–562. https://doi.org/10.1038/nmat2469.
(60) Zvezdin, A. K.; Mukhin, A. A. Magnetoelectric Interactions and Phase Transitions in a New Class of Multiferroics with Improper Electric Polarization. *JETP Lett.* **2008**, *88* (8), 505–510. https://doi.org/10.1134/S0021364008200083.
(61) Sahlot, P.; Jana, A.; Awasthi, A. M. Exchange Bias in Multiferroic Ca3Mn2O7 Effected by Dzyaloshinskii-Moriya Interaction. *AIP Conf. Proc.* **2018**, *1942* (1), 130009. https://doi.org/10.1063/1.5029079.
(62) JUNG, W.-H. Weak Ferromagnetism of n = 2 Ruddlesden : Popper Ca3Mn2O7 System. *Weak Ferromagn. N 2 Ruddlesden Popper Ca3Mn2O7 Syst.* **2000**, *19* (22), 2037–2038.
(63) Elcombe, M. M.; Kisi, E. H.; Hawkins, K. D.; White, T. J.; Goodman, P.; Matheson, S. Structure Determinations for Ca3Ti2O7, Ca4Ti3O10, Ca3.6Sr0.4Ti3O10 and a Refinement of Sr3Ti2O7. *Acta Crystallogr. B* **1991**, *47* (3), 305–314. https://doi.org/10.1107/S0108768190013416.
(64) Guiblin, N.; Grebille, D.; Leligny, H.; Martin, C. Ca3Mn2O7. *Acta Crystallogr. C* **2002**, *58* (1), i3–i5. https://doi.org/10.1107/S0108270101018492.
(65) Bendersky, L. A.; Chen, R.; Fawcett, I. D.; Greenblatt, M. TEM Study of the Electron-Doped Layered La2−2xCa1+2xMn2O7: Orthorhombic Phase in the 0.8. *J. Solid State Chem.* **2001**, *157* (2), 309–323. https://doi.org/10.1006/jssc.2000.9068.